\documentclass[12pt]{article}
\usepackage{graphicx,rotating}
\def\P{{\mathcal P}}
\def\R{{\mathcal R}}
\def\gm2{{\rm GeV}^{-2}}

\begin{document}

\rightline {LYCEN 2000-60}

\rightline {DFTT 28/2000}

\begin{center}
{\Large \bf Exchange-degenerate Regge trajectories:\\ a fresh look from
resonance and\\ forward scattering regions} \\
\end{center}

\vskip 0.5 cm
{\large \bf
P. Desgrolard $^{a,}$\footnote{E-mail: desgrolard@ipnl.in2p3.fr},
M. Giffon     $^{a,}$\footnote{E-mail: giffon@ipnl.in2p3.fr},
E. Martynov   $^{b,}$\footnote{E-mail: martynov@bitp.kiev.ua},
E. Predazzi   $^{c,}$\footnote{E-mail: predazzi@to.infn.it}
}

\bigskip

\noindent
$^a$ Institut de Physique Nucl\'eaire de Lyon, IN2P3-CNRS et
Universit\'e Claude Bernard, \\ 43 boulevard du 11 novembre 1918, F-69622
Villeurbanne Cedex, France

\noindent
$^b$ Bogoliubov Institute for Theoretical Physics, NAS of
 Ukraine, 43 Kiev-143, \\ Metrologicheskaja 14b, Ukraine

\noindent
$^c$ Dipartimento di Fisica Teorica, Universit\`a di Torino and Sezione INFN di
Torino, \\ I-10125 Torino, Italy

\bigskip
\begin{center}
{\large Abstract}\\
\bigskip

\begin{minipage}{14cm}

The exchange-degeneracy of the mesonic $f$, $\omega $, $\rho $ and
$a_{2}$ Regge trajectories, dominant at moderate and high
energies in hadron elastic scattering, is analyzed from two viewpoints.
The first one concerns the masses of the resonances lying on these
trajectories; the second one deals with the total cross-sections and
the ratios of the real to the imaginary parts of the forward amplitudes of
hadron and photon induced reactions. Neither set of data supports exact
exchange-degeneracy.
\end{minipage}
\end{center}

\newpage
%%%%%%%%%%%%%%%%%%%%%%%%%%%%%%%%%%%%%%%%
%%%%%%%%%%%%%%%%%%%%%%%%%%%%%%%%%%%%%%%%
%%%%%%%%%%%%%%%%%%%%%%%%%%%%%%%%%%%%%%%%
\section{Introduction}
A very convenient and useful method to group mesons and baryons in
families with definite quantum numbers, makes use of the so called
{\it Chew-Frautschi plot} (spin versus squared mass). It is a
graphic representation of Regge trajectories for given quantum
numbers. Early analyses of Regge trajectories hinted at remarkable
properties \cite{Coll}: they appeared to be essentially linear and
many of them coincide. The latter property came to be known as the
principle of {exchange-degeneracy (e-d) of Regge trajectories.

There are two kinds of exchange-degeneracy, qualified as {\it
strong} and {\it weak}. In weak exchange-degeneracy, only the
trajectories with different quantum numbers coincide. In strong
exchange-degeneracy, in addition, the residues of the
corresponding hadronic amplitudes coincide at the given pole in
the $j-$ plane. It was soon realized that strong
exchange-degeneracy may be violated (for theoretical arguments,
see \cite{Coll}) and indeed experimental confirmations of this
violation occurred.

Conclusive and definitive statements about weak exchange-degeneracy,
however, are not possible without a sufficiently precise
experimental information about the hadrons lying on each Regge
trajectory. Therefore, lacking high precision data, general
agreement with a weak exchange-degeneracy assumption, as well as
with a linearity of meson Regge trajectories, was claimed in the
past (see the references to old papers in \cite{PDGM}) and the
hypothesis was applied repeatedly, for example, in models
describing elastic scattering data (see references below). From
this point of view, the most relevant trajectories are the $f$, the
$\omega$, the $\rho$, and the $a_{2}$, which can variously be
exchanged in the $t$-channel of many elastic reactions. These we are
going to consider in what follows. The r\^ole of a unique Regge
trajectory was repeatedly analyzed to describe hadron-hadron and
photon-hadron total cross-sections in a most economical approach
\cite{DL}. In spite, however, of an apparent agreement with the
data, this model leads numerically to a quite large $\chi^{2}$
when compared with more recent approaches~\cite{DGLM,CEKLT}.

Today, the situation has changed somewhat. Three meson states are now
known lying on each trajectory (except for the $\omega-$tra\-jec\-to\-ry
for which we know only two states) and, moreover, some of their masses are
measured with very high precision \cite{PDGM} even though the data on
highest spin resonances have not yet been confirmed. We believe, however,
that a fairly conclusive analysis can be performed using, on the one hand
data in the resonance region (Section~2) and, on the other hand, data on
(near forward) elastic scattering (Section~3).

Our conclusion (Section~4), will suggest that the combined analysis of all
data supports a breaking of the weak exchange-degeneracy principle.
%%%%%%%%%%%%%%%%%%%%%%%%%%%%%%%%%%%%%%%%%%%%%%
%%%%%%%%%%%%%%%%%%%%%%%%%%%%%%%%%%%%%%%%%%%%%%
%%%%%%%%%%%%%%%%%%%%%%%%%%%%%%%%%%%%%%%%%%%%%%
\section{Resonance region}
To examine the agreement of weak exchange-degeneracy with the available
data in the resonance region, we first assume that the four trajectories
$f$, $\omega$, $\rho$, $a_{2}$ are linear and coincide.
Writing the relevant exchange-degenerate linear trajectory as
\begin{equation}\label{1}
  \alpha_{e-d}(m^{2})=\alpha_{e-d}(0)+\alpha'_{e-d}\ m^{2}\
\end{equation}
($m$ is the mass of the bound state),
we determine the intercept $\alpha_{e-d}(0)$ and the slope
$\alpha'_{e-d}$ by fitting 11 resonances lying on $f, \omega ,
\rho $ and $a_{2}$ trajectories. Using the MINUIT computer code,
we find (the precision is estimated as the usual one-standard
deviation error)
\begin{equation}\label{2}
\alpha_{e-d}(0)=0.4494\pm 0.0007, \quad \alpha'_{e-d}=(0.9013 \pm
0.0011) \ \gm2, \quad \mbox{with} \quad \chi^{2}/dof=117.9.
\end{equation}
The data are taken from Ref.~\cite{PDGM}. The very high value of
$\chi^{2}/dof$ ({\it dof} stands for degree of freedom defined as
the difference between the number of data points and the number of
 fitted parameters) is not surprising because {\it (i)} the data
exhibit a known nonlinearity of the trajectories (see details below)
and {\it(ii)} the
masses of the low lying resonances are measured with very high
precision.
The corresponding degenerate trajectory one obtains is shown in Fig.~1
(solid line). For
comparison, the trajectory with the parameters used in Ref.~\cite{GN},
$\alpha(m^{2})=0.48+0.88m^{2}$ ($m$ in GeV), is also plotted (dashed
line).

\vskip 1.cm

\begin{figure}[ht]
\begin{center}
\includegraphics[scale=0.70]{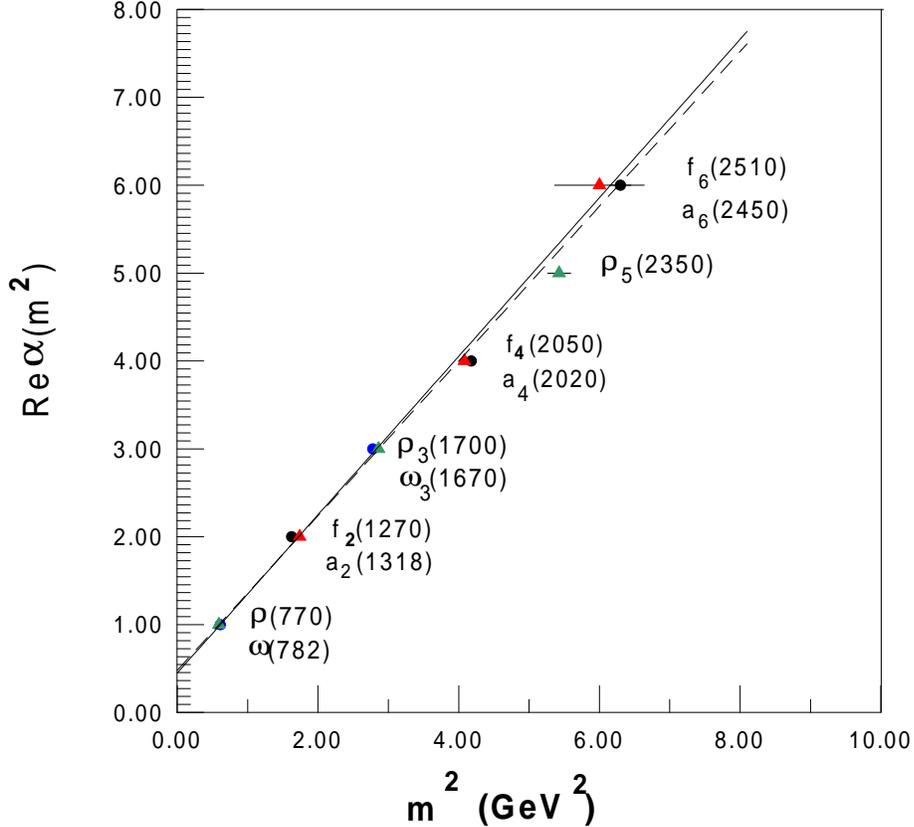}
\caption{Chew-Frautschi plot for the fully exchange-degenerate $f $,
$\omega $, $\rho $ and $a_2$ trajectories. The solid line denotes the
trajectory with the parameters obtained in our fit; the dashed line is the
trajectory from~\cite{GN}. }
\end{center}
\end{figure}
\medskip

Our conclusion is, thus, that in spite of a decent agreement with
resonance data (plotted \`a la Chew-Frautschi), weak exchange degeneracy
of the $f$, $\omega$, $\rho$, and $a_2$ trajectories is not supported by
the resonance data when a more precise numerical analysis is performed.

\smallskip

In order to verify the possibility of a limited validity of
exchange-degeneracy, we have considered a weaker (intermediate) version where
the trajectories are grouped in pairs. Some of these combinations
has been currently used to describe the behaviour of total
cross-sections and of their differences. For convenience, we present
the results in Table~1 where all the 6 possible groupings in pairs
are considered.
\begin{center}
\begin{tabular}{|c|c|c|c|c|}
\hline
          &  & $\omega$ & $\rho$ & $a_2$  \\
\hline
          & $\alpha(0)$      &  0.411   & 0.442  & 0.565  \\
   $f$    & $\alpha'(\gm2)$  &  0.963   & 0.944  & 0.835  \\
          & $\chi^{2}/dof$   &  66.84   & 57.34  & 194.26 \\
\hline
          & $\alpha(0) $     &          & 0.445  & 0.456  \\
 $\omega$ & $\alpha'(\gm2)$  &          & 0.908  & 0.890  \\
          & $\chi^{2}/dof$   &          & 84.14  & 13.48  \\
\hline
          & $\alpha(0)$      &          &        & 0.482  \\
  $\rho$  & $\alpha'(\gm2)$  &          &        & 0.874  \\
          & $\chi^{2}/dof$   &          &        &  1.30  \\
\hline
\end{tabular}
\end{center}
\noindent
Table~1. Intercepts $\alpha(0)$, slopes $\alpha'$ and
$\chi^{2}/dof$'s obtained in the fits when exchange-degeneracy is
assumed for each grouping in pairs of the trajectories. They are
written at the intersections of the corresponding line and row.

\medskip

For any grouping in pairs one can obtain the $\chi^{2}$ from the Table
because each pair is considered independently of the other. What
would appear as a {\it natural} grouping introducing just two
pairs of degenerate trajectories (one crossing even $f-a_{2}\equiv R_{+}$
and one crossing odd $\omega -\rho\equiv R_{-}$,
as in~\cite {PDG1}), is clearly not supported by the
resonance data under any reasonable {\it common $\chi^{2}$}.

An obvious general conclusion follows from this very simple analysis:
under a careful numerical investigation, there are no experimental
evidences from the resonance region that the $f, \omega, \rho$ and
$a_{2}$ trajectories can be assumed to be {\it exchange-degenerate}.

\bigskip

The available resonances are known with a good precision,
allowing the determination of intercept and slope of each trajectory taken
separately under the assumption of linearity
\begin{equation}\label{3}
\alpha_R(m^{2})=\alpha_R(0)+\alpha'_R. m^{2}\ ,\quad R=f,\omega,\rho,a_2\ .
\end{equation}

\begin{figure}
\begin{center}
\includegraphics*[scale=0.75]{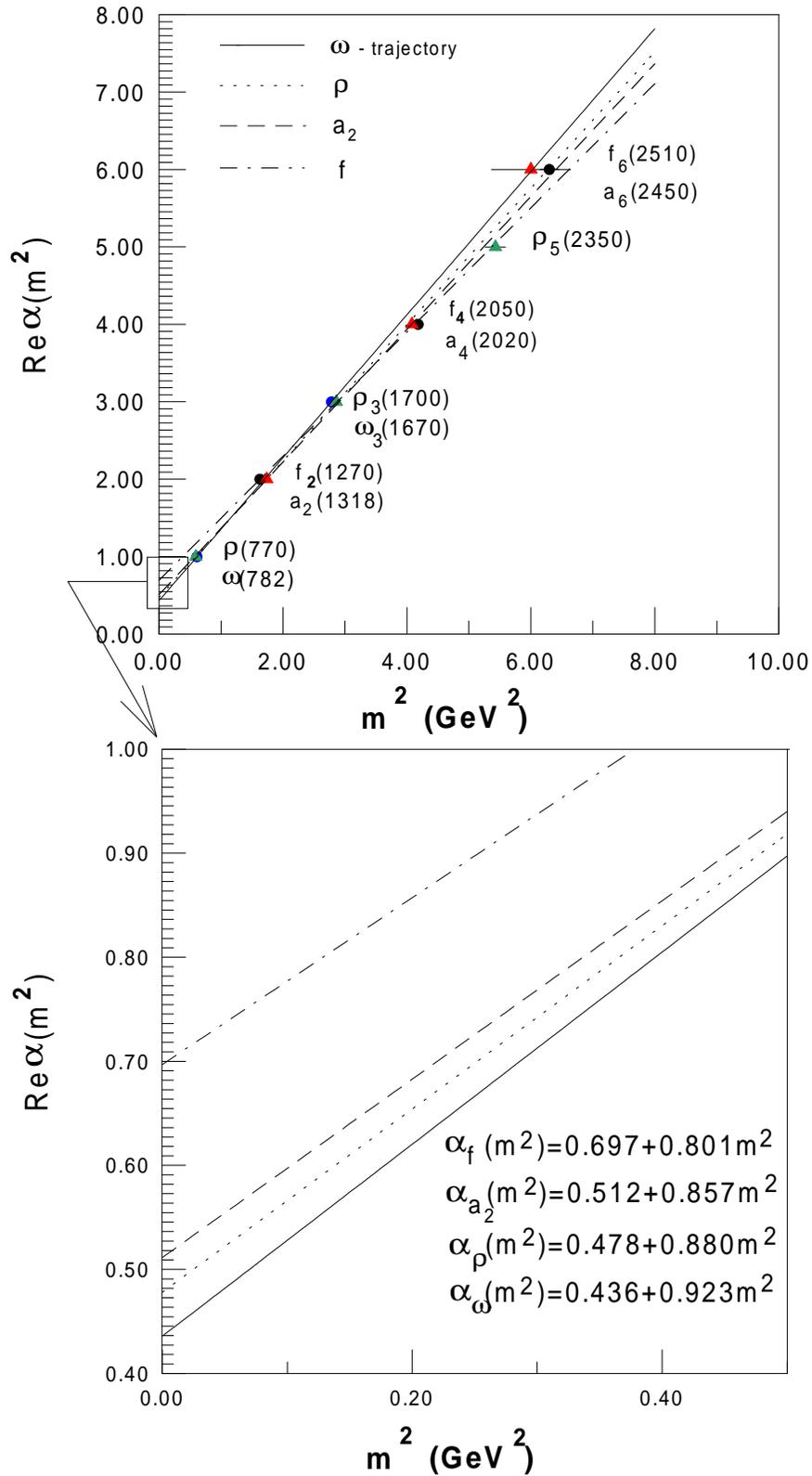}
\caption{Chew-Frautschi plots for $f$, $\omega$, $\rho$ and
$a_2$ Regge trajectories taken separately assuming linearity (the
figure below is an enlargement for low masses).}
\end{center}
\end{figure}
\smallskip

The corresponding Chew-Frautschi plots obtained from the fit are shown in
Fig.2. We obtained the following parameters
\begin{equation}\label{4}
  \begin{array}{lll}
    \alpha_{f}(0)=0.6971\pm0.0029, & \alpha'_{f}=(0.8014\pm0.0018)\ \gm2 ,
    & \chi^{2}/dof=6.01, \\
    \alpha_{\omega}(0)=0.4359, & \alpha'_{\omega}=0.9227\ \gm2 ,
    &   \mbox{(not fitted)},\\
    \alpha_{\rho}(0)=0.4783\pm0.0011, & \alpha'_{\rho}=(0.8800\pm0.0017)\ \gm2 ,
    & \chi^{2}/dof=3.31, \\
    \alpha_{a_{2}}(0)=0.5116\pm0.0009, & \alpha'_{a_{2}}=(0.8567\pm0.0008)\ \gm2 ,
    & \chi^{2}/dof=0.42\ , \
  \end{array}
\end{equation}
in qualitative agreement with the fits
of~\cite{CKK}. The rather high values 3 and 6 obtained for the $\chi^2$ of
two trajectories (which are anyhow much lower than in the assumption of
e-d) can be attributed to the hypothesis of linearity. Actually, we should
note that all trajectories, except the $\omega$ (for which only two
resonances are known), deviate from of a strict linear behaviour (see also
\cite{TanNor}).
Indeed, parametrizing in isolation the trajectories in a parabolic form instead
of a linear one
\begin{equation}\label{5}
\alpha_R(m^{2})=\alpha_R(0)+\alpha'_R. m^{2}+\frac{\alpha''_R}{2}m^{4} \ ,
\quad (R=f,\omega,\rho,a_2\,),
\end{equation}
one obtains from the experimental data on known resonances~\cite{PDGM} (two
for the $\omega$ and three for the other Reggeons)
\begin{equation}
\begin{array}{lll}\label{6}
f\ \mbox{\rm  trajectory:} &&\\
\alpha_f(0)=0.9577\pm 0.0023,&\alpha'_f=(0.5858\pm 0.0014) \mbox{\rm
GeV}^{-2},&\alpha''_f=(0.0681\pm 0.0015)\mbox{\rm GeV}^{-4},\\
\rho\ \mbox{\rm  trajectory:} &&\\
\alpha_\rho(0)=0.4404\pm 0.0011,&\alpha'_\rho=(0.9566\pm 0.0017)\mbox{\rm
GeV}^{-2},&\alpha''_\rho=(-0.0430\pm 0.0028)\mbox{\rm GeV}^{-4},\\
a_{2}\ \mbox{\rm  trajectory:} &&\\
\alpha_{a_{2}}(0)=0.8759\pm 0.0010,&\alpha'_{a_{2}}=(0.5987\pm 0.0006)
\mbox{\rm
GeV}^{-2},&\alpha''_{a_{2}}=(0.0876\pm 0.0007)\mbox{\rm GeV}^{-4}.
\end{array}
\end{equation}
Of course such a parametrization cannot be satisfactory from a
theoretical point of view (the negative sign of the second derivative of
the $\rho$ trajectory also is strange), it only suggests a
nonlinearity of the given trajectories. A more detailed investigation of
the phenomenon taking into account the actual widths of the resonances is
desirable.

The deviation from linearity, dictated both by analyticity and unitarity,
has often been discussed in the past. For a recent discussion on the
nonlinearity of the $f$ trajectory and its influence on the intercept,
see, in particular~\cite{mart,dgm}.
%%%%%%%%%%%%%%%%%%%%%%%%%%%%%%%%%%%%%%%%%%%%%
%%%%%%%%%%%%%%%%%%%%%%%%%%%%%%%%%%%%%%%%%%%%%
%%%%%%%%%%%%%%%%%%%%%%%%%%%%%%%%%%%%%%%%%%%%%
%\newpage
\section{Forward scattering}
%%%%%%%%%%%%%%%%%%%%%%%%%%%%%%%%%%%%%%%%%%%%%
%%%%%%%%%%%%%%%%%%%%%%%%%%%%%%%%%%%%%%%%%%%%%
\subsection{Generalities}

The exchange-degeneracy hypothesis for the $f$, $\omega$, $\rho$,
$a_{2}$ trajectories can be checked also using elastic hadron
scattering data. In particular, one can use forward scattering
data, {\it i.e.} the total cross-sections for hadron hadron,
$\gamma$ hadron and $\gamma \gamma$-collisions. Following
the arguments given in \cite{CEKLT} we do not restrict our analysis to the
data on total cross-sections, $\sigma^{(t)}(s))$, but include in the fits the
ratios ($\rho$) of the real to the imaginary parts of the forward
amplitudes.

Performing such an analysis requires an explicit parametrization
for the amplitudes of the processes under investigation~\footnote{The best
way to analyze exchange-degeneracy would be to consider some linear
combinations of $\sigma^{(t)}$ for several elastic processes. In
principle, one can construct combinations that contain the contribution of
one or two Reggeons and these could be compared with the experiment.
The shortcoming of this procedure, however, lies in the fact that,
usually, the required reactions are measured at different energies.
As a consequence, while attractive, this procedure is heavily affected
by the ambiguity of reconstructing data from interpolation. We
shall not use this method.}. Like in the resonance-case, in order to
check  how well the exchange-degeneracy hypothesis works in the
description of hadron and photon induced  cross-sections, it is
not sufficient to obtain agreement with the data which {\it looks
good}. It is also necessary to compare this description with the one
where the e-d assumption is removed. Clearly, removing the assumption of
exchange degeneracy increases the number of parameters but the
$\chi^2$ referred to the number of degrees of freedom retains its
comparative validity.
Thus, we analyze the data using the following explicit expressions
for the forward amplitudes $A_{ab}(s,t=0)$ of the 12 elastic reactions
\begin{equation}\label{7}
  \begin{array}{lcl}
    A_{p^{\pm}p}(s,0)&= & \P_{NN}(s)+f_{NN}(s)+a_{NN}(s)\mp\omega_{NN}(s)
    \mp\rho_{NN}(s), \\
    A_{p^{\pm}n}(s,0)&= & \P_{NN}(s)+f_{NN}(s)-a_{NN}(s)\mp\omega_{NN}(s)
    \pm\rho_{NN}(s),  \\
    A_{\pi^{\pm}p}(s,0)&= & \P_{\pi N}(s)+f_{\pi N}(s)\mp \rho_{\pi N}(s), \\
    A_{K^{\pm}p}(s,0)&= & \P_{KN}(s)+f_{KN}(s)+a_{KN}(s)\mp\omega_{KN}(s)
    \mp\rho_{KN}(s), \\
    A_{K^{\pm}n}(s,0)&= & \P_{KN}(s)+f_{KN}(s)-a_{KN}(s)\mp\omega_{KN}(s)
    \pm\rho_{KN}(s), \\
    A_{\gamma p}(s,0)&= & \delta\P_{NN}(s)+f_{\gamma N}(s), \\
    A_{\gamma \gamma}(s,0)&= & \delta^{2}\P_{NN}(s)+f_{\gamma \gamma N}(s) \
    ,\\
\end{array}
\end{equation}
($p^+,p^-$ stand for $p,\bar p$).

In addition to the main goal (to compare the e-d hypothesis with
the data), we test two models of Pomeron, each
one with two components: a constant background and an energy dependent term.
One of them, explored in \cite{GN,DN}, is
{\it universal}, in the sense that its asymptotic component,
growing with energy, contributes equivalently to all processes
\begin{equation}\label{8}
  \P_{ab}(s)=i\{Z_{ab}+X\P(s)\}\qquad \mbox{for "universal" Pomeron}\ .
\end{equation}
The other one is {\it non-universal}: its two components
contribute differently to each process, but with a universal
ratio of these two components.
\begin{equation}\label{9}
  \P_{ab}(s)=iZ_{ab}\{X+\P(s)\}\qquad \mbox{for "nonuniversal" Pomeron}\ .
\end{equation}

 We remark that the suggestion to
consider models with a "two-component" Pomeron is not new. Many
times, this idea was successfully applied (see~\cite{DGLM,maqm}
and references therein).
Different constant terms $Z_{ab}$ in (8) or $Z_{ab}X$ in (9), often
neglected, are intended to adjust the universal behavior of a unique Regge
term
and will reveal to improve the $t=0$ fits of a simple one-component Pomeron
especially at medium-high energies.
>From the phenomenological point of view, an energy-rising Pomeron
component is only an asymptotic part of its contribution,
unknown subasymptotic terms must also exist contributing to the amplitudes. We
take into account effectively this part of Pomeron when adding a constant term
to $A(s,0)$. It corresponds to a simple $j$-pole with a unit intercept.
Various theoretical justifications of the existence of such an additive
structure of the Pomeron (or of the total cross section) have been
proposed : we note that some indication for such a background
component has been found recently along with the ordinary BFKL Pomeron
\cite{lip} (in contrast to the known hard component produced by two-gluon
states, the new found one is constructed from the  three-gluon states but
with positive $C$-parity differing, from the three-gluonic Odderon with
negative $C$-parity). At an equally fundamental level~\cite{KPPP}, such a
constant term has been recognized as the nonperturbative
contribution that one must add to the perturbative soft gluon radiation
term responsible for the growth with energy of total hadronic cross
sections.

We have considered two variants for the $s-$depen\-dent Pomeron component,
having in mind its properties in the complex angular momentum plane. The
first one corresponds to a simple pole in the complex angular momentum
plane with intercept $\alpha_{\P}(0)=1+\epsilon$, (the so-called {\it
Supercritical Pomeron} (SCP))
\begin{equation}\label{10}
  \P(s)=(-is/s_{0})^{\epsilon}, \qquad s_{0}=1\mbox{GeV}^{2}.
\end{equation}
The second variant corresponds to the {\it Dipole Pomeron} (DP). In the
$j$-plane it is described by a double pole with a unit intercept
trajectory,
$\alpha_{\P}(0)=1$
\begin{equation}\label{11}
  \P(s)=\ln(-is/s_{0}).
\end{equation}

\smallskip
For the secondary Reggeons we use the standard form
\begin{equation}\label{12}
  \R _{ab}(s)=\eta Y_{\R ab}(-is/s_{0})^{\alpha_{\R}(0)-1}\, \quad
  R=f, a_{2}, \omega, \rho\ ,
\end{equation}
where $\eta=i$ for $f$ and $a_{2}$ while $\eta=1$ for
$\omega$ and $\rho$.

\smallskip
The above amplitudes are normalized according to
\begin{equation}\label{13}
  \sigma_{ab}^{(t)}(s)=8\pi \Im mA_{ab}(s,0).
\end{equation}
%%%%%%%%%%%%%%%%%%%%%%%%%%%%%%%%%%%%%%%%%%%%%%%%%%%%%%%%%
%%%%%%%%%%%%%%%%%%%%%%%%%%%%%%%%%%%%%%%%%%%%%%%%%%%%%%%%%
%%%%%%%%%%%%%%%%%%%%%%%%%%%%%%%%%%%%%%%%%%%%%%%%%%%%%%%%%
\subsection{Results}
We have taken into account the whole set of cross-section data for
$(p^{\pm}p)$, $(p^{\pm}n)$, $(K^{\pm}p)$, $(K^{\pm}n)$, $(\pi^{\pm}p)$,
$(\gamma p)$ and
$(\gamma \gamma )$ interactions and of $\rho$ ratio data for all interactions
excluding two last. Furthermore, in order to reasonably
neglect the sub-leading meson trajectories,
and to respect the stability if the  $\chi^2$ and of the parameters (see the
discussion below), we choose the energy range with $\sqrt{s}\ge 5$ GeV. No
other {\it wise selection} of any kind is attempted (such as a filtering
of the data suggested by some authors). In total there are 785 points
available in the Data Base of Particle Data Group \cite{PDG1}.

The values of the fitted parameters are given in Table~2 for
the universal and the non universal Pomeron. If exchange-degeneracy is assumed,
all intercepts of $f$, $a_{2}$, $\omega$ and $\rho$ Reggeons are
equal (in Table 2, we have labeled the common intercept
as $\alpha_{f}(0)$). We do not give the errors for the other parameters than
intercepts, neither curves for the total
cross-sections and $\rho$ ratios  because they are only illustrative,
not very important for the case in point.

One can see that in all considered cases, non degenerate
trajectories lead to a better $\chi^2$, even though for
the universal Supercritical Pomeron the difference is very small.
The best agreement with the fitted data ("measured" by the $\chi^2$) is
obtained either with the non universal non-degenerate DP or the SCP
(compare columns 8 and 6 in Table~2). The reason lies in the
similarity of these models when (as in the present case), the value of
$\epsilon =\alpha_\P(0)-1$ is very small ($\epsilon \approx 0.0101 $).
Actually, as emphasized in \cite{maqm,DGLM}, when $\epsilon\ll 1$, the
Supercritical Pomeron approximates very closely the Dipole Pomeron since,
in the relevant energy range, and for $s_0$ given in (10)
$$
Z_{ab}\left[X+(-is/s_{0})^{\epsilon}\right]\approx Z_{ab}\left[1+X+\epsilon
\ell n(-i{s\over s_0})\right]\ \equiv Z'_{ab}\left[ X'+\ell n(-{is\over s_0})
\right],
$$
and this reflects in the numerical values found in Table 2.

%\hspace{+2.2cm}
\begin{sideways}
{\scriptsize
 $
\begin{array}{|c|c|c|c|c||c|c|c|c|}
\hline
                               & \multicolumn{4}{c||}{\rm \bf Universal \, \, Pomeron}                         & \multicolumn{4}{c|}{\rm \bf Nonuniversal \, \, Pomeron} \\
\cline{2-9}
                               & \multicolumn{2}{c|}{\rm Dipole}   & \multicolumn{2}{c||}{\rm Supercritical}   & \multicolumn{2}{c|}{\rm Dipole}   & \multicolumn{2}{c|}{\rm Supercritical}\\
\cline{2-9}
                                       & \rm n e-d  & \rm e-d      & \rm n e-d    & \rm e-d                    & \rm n e-d     & \rm e-d      & \rm n e-d    & \rm e-d  \\
\hline
       \chi^2/dof                      & .1215E+01  & .1886E+01    & .9987E+00    & .1089E+01                   & .9575E+00     & .1717E+01    & .9585E+00    &  .1349E+01     \\
\hline
        \epsilon                       & .0000E+00  & .0000E+00    & .1402E+00    & .1757E+00                   & .0000E+00     & .0000E+00    & .1013E-02    &  .1267E+00     \\
     \Delta_{\epsilon}                 &            &              & .6493E-02    & .5059E-02                   &               &              & .3861E-06    &  .1831E-02     \\
       Z_{NN}, \, {\rm GeV}^{-2}       &-.2651E+01  & .2054E+01    & .1962E+01    & .2790E+01                   & .6756E+00     & .3221E+00    & .6480E+03    &  .8931E+00     \\
    Z_{\pi N}, \, {\rm GeV}^{-2}       &-.3826E+01  & .5210E+00    & .4984E+00    & .1254E+01                   & .4596E+00     & .1986E+00    & .4394E+03    &  .5495E+00     \\
       Z_{KN}, \, {\rm GeV}^{-2}       &-.3950E+01  & .2084E+00    & .2077E+00    & .9321E+00                   & .4342E+00     & .1738E+00    & .4139E+03    &  .4789E+00     \\
         X, \, {\rm GeV}^{-2}          & .6222E+00  & .2708E+00    & .7396E+00    & .3845E+00                   &-.4380E+01     & .5195E+01    &-.1004E+01    &  .2132E+01     \\
\hline
     \alpha_{f}(0)                     & .8063E+00  & .5481E+00    & .5926E+00    & .4904E+00                   & .7895E+00     & .5396E+00    & .7850E+00    &  .5002E+00     \\
     \Delta_{f}                        & .3343E-03  & .2383E-03    & .1225E-01    & .5931E-02                   & .5330E-03     & .3943E-02    & .1333E-02    &  .6827E-02     \\
      Y_{fNN}, \, {\rm GeV}^{-2}       & .1021E+02  & .9363E+01    & .7037E+01    & .8623E+01                   & .1128E+02     & .1130E+02    & .1110E+02    &  .1002E+02     \\
     Y_{f\pi N}, \, {\rm GeV}^{-2}     & .8625E+01  & .6785E+01    & .4399E+01    & .4997E+01                   & .6461E+01     & .5296E+01    & .6306E+01    &  .4113E+01     \\
      Y_{fKN}, \, {\rm GeV}^{-2}       & .7745E+01  & .5184E+01    & .2978E+01    & .2863E+01                   & .5169E+01     & .2870E+01    & .5003E+01    &  .1462E+01     \\
   Y_{f\gamma N}, \, {\rm GeV}^{-2}    & .2795E-01  & .1762E-01    & .1095E-01    & .1127E-01                   & .3100E-01     & .2234E-01    & .3018E-01    &  .1556E-01     \\
 Y_{f\gamma \gamma}, \, {\rm GeV}^{-2} & .7412E-04  & .2598E-04    & .5349E-05    &\approx 0                    & .8179E-04     & .3476E-04    & .7848E-04    &  .9098E-05     \\
\hline
    \alpha_{\omega}(0)                 & .4842E+00  &=\alpha_{f}(0)& .4518E+00    &=\alpha_{f}(0)               & .4459E+00     &=\alpha_{f}(0)& .4447E+00    &=\alpha_{f}(0)  \\
      \Delta_\omega                    & .7427E-02  &= \Delta_{f}  & .1048E-01    &=\Delta_{f}                  & .9905E-02     & \Delta_{f}   & .9781E-02    &=\Delta_{f}     \\
   Y_{\omega NN}, \, {\rm GeV}^{-2}    & .3866E+01  & .3179E+01    & .4284E+01    & .3779E+01                   & .4407E+01     & .3272E+01    & .4425E+01    &  .3679E+01     \\
   Y_{\omega KN}, \, {\rm GeV}^{-2}    & .1295E+01  & .1089E+01    & .1418E+01    & .1265E+01                   & .1438E+01     & .1100E+01    & .1443E+01    &  .1218E+01     \\
\hline
    \alpha_{\rho}(0)                   & .5131E+00  &=\alpha_{f}(0)& .5299E+00    &=\alpha_{f}(0)               & .5115E+00     &=\alpha_{f}(0)& .5123E+00    &  =\alpha_{f}(0) \\
      \Delta_\rho                      & .1967E-01  &=\Delta_{f}   & .2017E-01    &=\Delta_{f}                  & .2158E-01     &  \Delta_{f}  & .2156E-01    &  \Delta_{f}    \\
    Y_{\rho NN}, \, {\rm GeV}^{-2}     & .1654E+00  & .1557E+00    & .1596E+00    & .1813E+00                   & .2045E+00     & .1962E+00    & .2043E+00    &  .2114E+00    \\
   Y_{\rho \pi N}, \, {\rm GeV}^{-2}   & .6950E+00  & .6629E+00    & .6685E+00    & .7344E+00                   & .6833E+00     & .6523E+00    & .6822E+00    &  .6996E+00      \\
    Y_{\rho KN}, \, {\rm GeV}^{-2}     & .3691E+00  & .3383E+00    & .3428E+00    & .3822E+00                   & .3589E+00     & .3332E+00    & .3578E+00    &  .3619E+00      \\
\hline
    \alpha_{a_2}(0)                    & .7918E+00  &=\alpha_{f}(0)& .6933E+00    & =\alpha_{f}(0)              & .6912E+00     &=\alpha_{f}(0)& .6868E+00    &=\alpha_{f}(0)    \\
       \Delta_{a_{2}}                  & .9249E-01  &=\Delta_{f}   & .1081E+00    &=\Delta_{f}                  & .2343E-01     & \Delta_{f}   & .1189E+00    &   \Delta_{f}     \\
     Y_{a_2 NN}, \, {\rm GeV}^{-2}     & \approx 0  & \approx 0    & \approx 0    & \approx 0                   &\approx 0      &\approx 0     &\approx 0     & \approx 0      \\
     Y_{a_2 KN}, \, {\rm GeV}^{-2}     & .4082E-01  & .1571E+00    & .6986E-01    & .2148E+00                   & .6897E-01     & .1588E+00    & .7069E-01    &  .1974E+00       \\
\hline
    \delta                             & .3427E-02  & .3031E-02    & .3096E-02    & .3019E-02                   & .3453E-02     & .3056E-02    & .3440E-02    &  .3043E-02       \\
\hline
\end{array}
$
}
\end{sideways}

\medskip
\noindent
Table 2. Results of the fits to forward elastic data using the universal
and non universal Pomeron showing  the differences  when
exchange-degeneracy is assumed (e-d) and not assumed (ne-d). Dipole and
Supercritical Pomeron are worked out. The estimated one-standard deviation
errors are reported only for $\epsilon=\alpha_{P}(0)-1$ and for
the four Reggeons intercepts $\alpha_{R}(0)$;
they are are denoted as $\Delta_\epsilon$ and $\Delta_R$.

\smallskip

The authors of \cite{CEKLT} rightly insist that every model should be
verified for the stability of its parameters and $\chi^2$ under change of
$\sqrt{s_{min}}$ (the minimal energy for the set of experimental data used
in the fit). For example, they have found that the Supercritical Pomeron
model and the model with $\sigma^{(t)}\propto \ell n^{2}s$ are stable for
$\sqrt{s}\geq \sqrt{s_{min}}  =9$ GeV, while the Dipole Pomeron model is
stable for $\sqrt{s}\geq \sqrt{s_{min}}  =5$ GeV. We agree, in general,
with this comment but, in our case, it is essential to keep $s_{min}$ as
small as possible, because the contribution of the secondary Reggeons
decreases with energy. Increasing $s_{min}$, we loose the data which
allow to discriminate between the different Reggeons and to test the e-d
assumption.

To check our conclusions, nevertheless, we have performed such a stability
analysis. In Figs.~3-4, the results are reported for the meaningful cases
of non degenerate trajectories. We show the dependence of the four Reggeon
intercepts and of the $\chi^{2}$'s versus $\sqrt{s_{min}}$ for two
representative models already discussed. The first is the Dipole Pomeron
model with a non universal Pomeron term (Fig.~3) (as  explained
previously, due to the smallness of $\epsilon$, also the non universal SCP
model is well approximated by this non universal DP model).
The second one is the Supercritical Pomeron Model with a universal form of
Pomeron (Fig.~4).
\smallskip

\begin{figure}[ht]
%%\begin{minipage}[t]{7.1cm}
\begin{center}
\includegraphics*[scale=0.60]{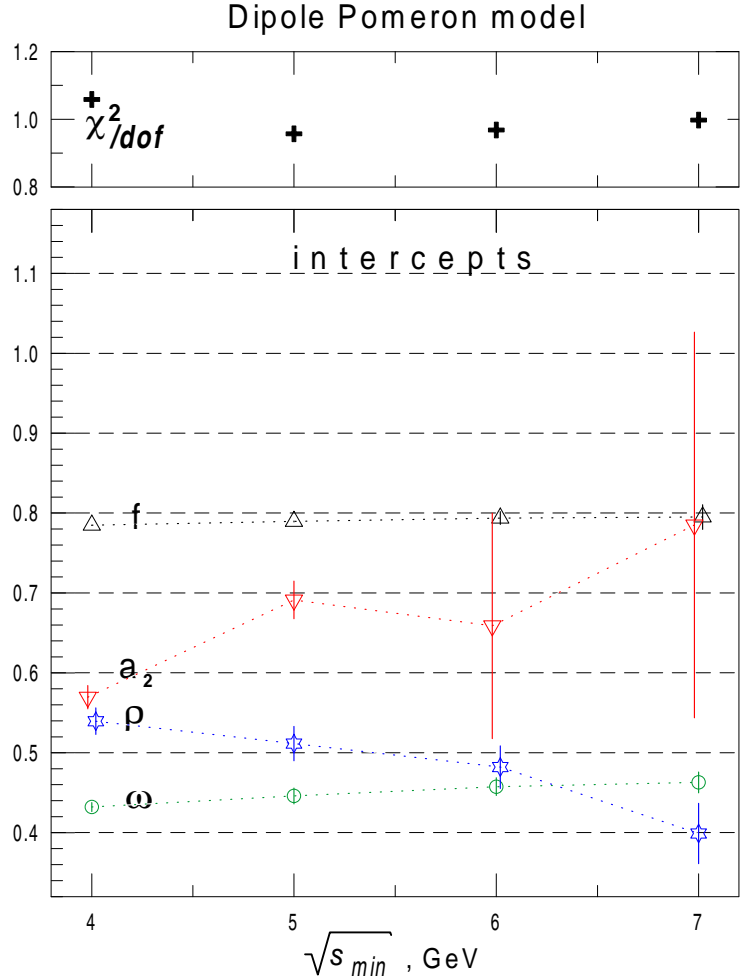}
\caption{Stability of the $\chi^2$ (upper part) and of the
Reggeons intercepts (lower part) versus the minimal c.m. energy limiting
the fitted data, for the non degenerate non universal Dipole Pomeron
model. The dashed lines join the central values for each Reggeon for
visual indication. Some points are shifted slightly to the left (right)
side to make the errors more easily distinguishable.}
\end{center}
%%\end{minipage}
%%\hskip .5cm
%%\begin{minipage}[t]{7.1cm}
\end{figure}
\begin{figure}[ht]
\begin{center}
\includegraphics*[scale=0.60]{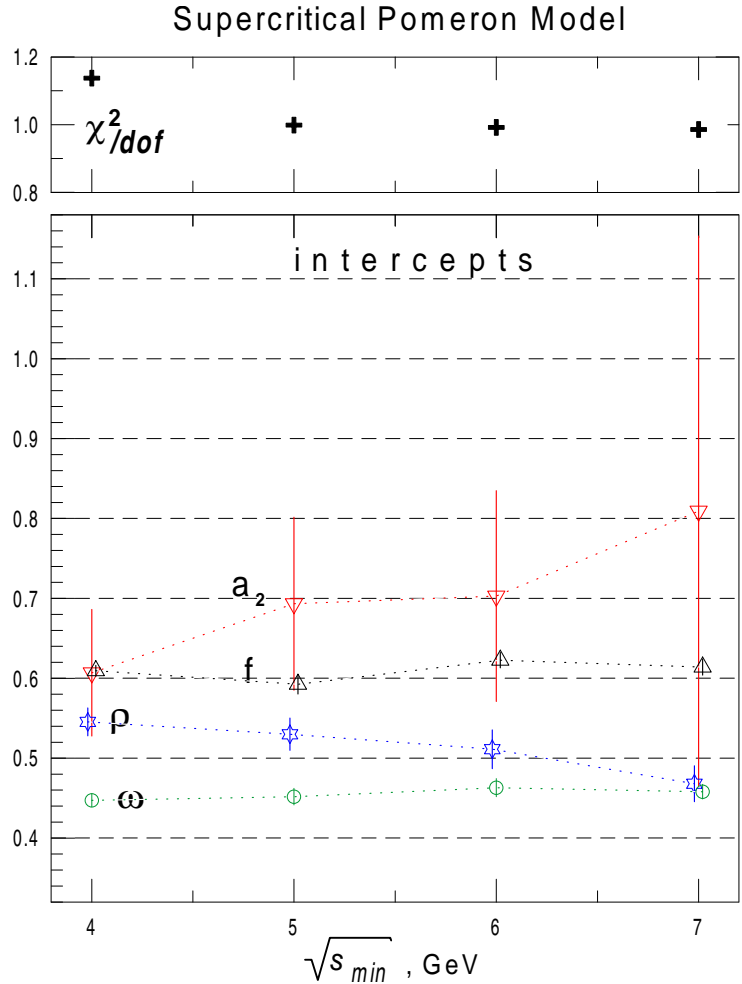}
\caption{ Same as in Fig.~3 for the non degenerate universal Supercricical
Pomeron model.}
\end{center}
%%\end{minipage}
\end{figure}
\medskip

One can see from these figures that the errors on the $\rho$ and,
especially on the $a_{2}$ trajectories increase with $s_{min}$. This was
expected since the $a_{2}$ contribution is determined mainly from rather
poor $(pn)$ and $(Kn)$ data. The situation with the $\rho$ contribution is
better, due to the available $(\pi p)$ data. In general, we see that the
scattering models used here are quite stable in the region $5$ GeV $\leq
\sqrt{s_{min}}\leq 7$ GeV; this, in itself, is a justification of our $5$
GeV minimal choice.

To complete this study, we performed also fits of the non universal Dipole
Pomeron and universal Supercritical Pomeron with non degenerate Reggeons
fixing their intercepts at the values determined in the Section 2 from the
resonance data (6). For DP we obtained $\chi^{2}/dof=1.034$ while for
SCP $\chi^{2}/dof=1.055$. Thus, no contradiction appears
between the forward scattering and the spectroscopy data.

From this analysis of forward scattering data, we can argue that, taking
into account the values of the intercepts $(\alpha(0))$, the errors in
their determination ($\Delta$), and the $\chi^{2}$'s (Table~2) that the
solution with nondegenerate Regge trajectories is definitely to be
preferred.

\medskip
%%%%%%%%%%%%%%%%%%%%%%%%%%%%%%%%%%%%%%%%%%%%%%%%%%%%%%
%%%%%%%%%%%%%%%%%%%%%%%%%%%%%%%%%%%%%%%%%%%%%%%%%%%%%%
%%%%%%%%%%%%%%%%%%%%%%%%%%%%%%%%%%%%%%%%%%%%%%%%%%%%%%
%\newpage
\section{Conclusions.}
In the first part of this work, we have concluded that the assumption
of exchange degeneracy of the $f$, $\omega$, $\rho$ and $a_{2}$ Regge
trajectories (assumed to be linear), though qualitatively acceptable, is
not compatible with a numerical best fit of the available data on the
corresponding mesonic resonances.

Concerning the forward scattering data, considered in the second part, the
situation is less clear because a reasonable description of the $t=0$ data
can be obtained under both hypotheses and is {\it a priori} model
dependent. We have tried to eliminate or at least to weaken this model
dependence in our conclusion by analizing four models that provide a
good description of forward scattering data. The fits with non degenerate
trajectories invariably improve the $\chi^{2}$'s, (all best fits occur for
non degenerate parametrizations), and this can be taken as an indication
in favor of the non degeneracy assumption.

For a more definitive conclusion we would need more precise data on
meson-nucleon and proton-neutron or K-neutron cross-sections at higher
energies. Even more conclusive, perhaps, would be to compare fits with and
without exchange degeneracy involving {\it all} data both at $t=0$ and at
$t \ne 0$. This analysis puts much more stringent constraints on the free
parameters as we have learned in previous experiences.

Thus, given that any model for scattering amplitudes should be in
agreement with both types of data, from spectroscopy and from total
cross-sections we conclude that the hypothesis of {\it exact
exchange degeneracy}, even in its weak formulation, is not supported by
the present data. In spite of this, due to its great economy in the number
of parameters, exchange degeneracy associated  with linear Regge trajectories
retains its usefulness in practical
calculations when only a rough approximation is sufficient.

\medskip
{\bf Acknowledgments} E.M. would like to thank the Universit\'e
Claude Bernard and IPNL of France and the MURST and INFN of Italy for the
kind hospitality and financial support during this work.

\bigskip

%%%%%%%%%%%%%%%%%%%%%%%%%%%%%%%%%%%%%%%%%%%%%%%%%%%%%%%%%%%%%%%%%%%%%%%%%%%%%
%%%%%%%%%%%%%%%%%%%%%%%%%%%%%%%%%%%%%%%%%%%%%%%%%%%%%%%%%%%%%%%%%%%%%%%%%%%%%
%%%%%%%%%%%%%%%%%%%%%%%%%%%%%%%%%%%%%%%%%%%%%%%%%%%%%%%%%%%%%%%%%%%%%%%%%%%%%

\end{document}